\documentclass[12pt]{iopart}

\usepackage{amssymb}
\usepackage{graphicx}
\usepackage{placeins}
\usepackage{float}
\usepackage{subfig}
\usepackage{color}
\usepackage{float}
\usepackage{caption}
\usepackage{hyperref}
\newcommand{\beq}{\begin{equation}}
\newcommand{\ene}{\end{equation}}
\newcommand{\bull}{\hfill $\Box$}

\begin{document}

\title[Structural frequencies]{Membrane-in-the-middle optomechanical system and structural frequencies}

\author{Luis Pedro Lara$^{1}$, Ricardo Weder$^{2}$, and Luis Octavio Casta\~{n}os-Cervantes$^{3}$}

\address{$^{1}$ Instituto de F\'{\i}sica Rosario, CONICET-UNR, Blvd. 27 de Febrero 210 bis, 2000, Rosario, Santa F\'e, Argentina}
\address{$^{2}$ Departamento de F\'{i}sica Matem\'{a}tica, Instituto de Investigaciones en Matem\'{a}ticas Aplicadas y en Sistemas, Universidad Nacional Aut\'{o}noma de M\'{e}xico, Apartado Postal 20-126, Ciudad de M\'{e}xico  01000, M\'{e}xico.}
\address{$^{3}$ Instituto Tecnol\'{o}gico y de Estudios Superiores de Monterrey, Escuela de Ingenier\'{\i}a y Ciencias, 14380 Ciudad de M\'exico, CDMX , M\'{e}xico}
\ead{\mailto{lplara2014@gmail.com}, \mailto{weder@unam.mx}, \mailto{luis.castanos@tec.mx}}

\hspace{1.5cm} Correspondig author: Ricardo Weder
\begin{abstract}
We consider a one-dimensional {\it  membrane-in-the-middle}  model for a  cavity that consists of two fixed, perfect mirrors and a mobile dielectric membrane between them that has a constant electric susceptibility. We present a sequence of exact cavity angular frequencies that we call {\it structural angular frequencies} and that have the remarkable property that they are independent of the position of the membrane inside the cavity. Furthermore, the case of a thin membrane is considered and \textcolor{black}{simple, approximate formulae} for the  angular frequencies and for the modes of the cavity are obtained. \textcolor{black}{Finally, the cavity electromagnetic potential is numerically calculated and it is found that the potential is accurately described by a multiple scales solution.}   
\end{abstract}

\noindent
\submitto{\jpa}
\maketitle

\section{Introduction}

Cavity optomechanics investigates the interaction between radiation fields and mechanical oscillators \cite{Marquardt.2}. It has enabled the study of mesoscopic and macroscopic systems in the quantum regime and has also proved to be useful to study very rich and complex classical, nonlinear dynamics \cite{Thompson}-\cite{Fig2020} and its relationship to the entanglement dynamics between the mechanical oscillator and the field \cite{CND}. Moreover, it has found numerous applications such as precision metrology and quantum and classical communication \cite{Marquardt.2} and gravitational wave detection \cite{gra1,gra2}. Further, the mechanical oscillators can be coupled to radiation fields in both the optical and microwave regimes, so they can be implemented as mechanical microwave-optical converters \cite{Converter,MIM2} and have the potential of becoming a link between devices working in different frequency regimes. In fact, cavity optomechanics in the microwave regime has emerged as an area that has the potential to develop new classical electronic devices that can even operate at room temperature \cite{Room,Room2}. Especially relevant for the study, design, and optimization of such devices,  a classical electric circuit modeling microwave optomechanics has been developed \cite{circuit} and it has been successfully tested in an optomechanical system working in the classical regime without fitting parameters \cite{MST}. Comparison of this classical electric circuit model with the standard quantum treatment permits the establishment of bounds between the quantum and classical regimes and allows the identification of truly quantum features in optomechanical systems \cite{circuit}. This is especially relevant because an analysis of the nonclassicality of optomechanical phases has revealed that many effects can be reproduced classically \cite{QCP}.

Cavity optomechanics uses the dependence of the cavity resonance frequency on the position of the mechanical oscillator to couple the electromagnetic field with the mechanical oscillator \cite{Marquardt.2}. In addition to this dispersive coupling, one must include both  dissipative and  coherent couplings to complete the set of optomechanical interactions \cite{Li2019}. The former arises because  the cavity decay rate depends on the position of the mechanical oscillator and the latter corresponds to the coupling between cavity modes due to the motion of the mechanical oscillator. In particular, the dissipative coupling can replace the dispersive coupling to perform many tasks such as optomechanical cooling and mechanical sensing, see \cite{MOUT} and references therein. Typically, the dispersive coupling dominates over the dissipative coupling, so experiments that use the dissipative coupling to perform the aforementioned tasks usually require specific tuning conditions or setups where the dispersive coupling vanishes. Among the several experimental setups that use the dissipative coupling, see \cite{MOUT} and references therein, the \textit{membrane-in-the-middle} (MIM) optomechanical setup can present a large dissipative coupling and has the advantage that the resonator and mechanical oscillator can be independently optimized. The latter requires a complex positioning setup that has limited the cavity size and the scalability of such systems, so studies have been carried out to overcome this difficulty and have resulted in similar systems where the membrane is displaced from the middle of the cavity \cite{MIM1}. Moreover, it has been shown that such setups can greatly reduce radiation force noise without affecting the coupling between the membrane and the field \cite{QDC2}. Actually, the MIM optomechanical set up  combines good properties of the cavity, namely, highly reflective mirrors, with good properties of the mechanical oscillator, in particular,  very thin, light membranes that can be moved easily by radiation pressure.

In this article we consider the MIM model studied in \cite{Law, 2016} where the membrane is a slab with constant electric susceptibility. \textcolor{black}{The paper differs greatly from \cite{Law, 2016} because the objective is completely different. Reference \cite{Law} sought to establish a nonadiabatic quantum Hamiltonian for the system, while \cite{2016} wanted to determine how the classic electromagnetic field of system evolves. Here, the objective consists in characterizing some properties of the cavity resonances. The main, new result is that we establish a sequence of {\it structural angular frequencies} that have the remarkable property that they are independent of the position of the membrane inside the cavity.} As such, these could lead to new physics in these types of systems, since there would be no dispersive coupling. Furthermore, \textcolor{black}{another new result is that we} consider the case of a thin membrane and obtain approximate analytic formulae with an error term for the  angular frequencies and the modes of the cavity. Finally, we show that the potential obtained by solving the appropriate wave equation is \textcolor{black}{accurately described by a multiple scales solution}. The  MIM model that we consider is a simplification of the models that are used for theoretical studies and in experiments, and it contains some of the essential features of these models. We provide rigorous mathematical results on the angular frequencies, i.e., the resonances of our model, that shed a light in what can happen in experimental setups.
 We impose a law to the movement of the membrane to simulate the interaction with the electromagnetic field. Note that one dimensional models of a cavity with a membrane that moves according to a law that is given by an external agent are used in other contexts, for example, in the dynamical Casimir effect \cite{Dodonov}. 

The article is organized as follows. In Sec.~\ref{model} we present the  MIM model and recall some results from  \cite{2016}. \textcolor{black}{The following sections present all the new results.} In Sec.~\ref{slab} we \textcolor{black}{deduce} the {\it structural angular frequencies}, the approximate cavity angular frequencies \textcolor{black}{and modes} for a thin membrane, and compare the potential obtained numerically with the one given by a multiple scales approximation. Finally, in Sec.~\ref{conclusions} we give our conclusions.

\section{The membrane-in-the- middle model}\label{model}

In this section we briefly introduce the {\it membrane-in-the-middle} (MIM) model used in \cite{Law,2016}. It was developed to help gain insight into the physics of a mobile membrane interacting with a cavity field.

\begin{figure}
\includegraphics[width=8cm]{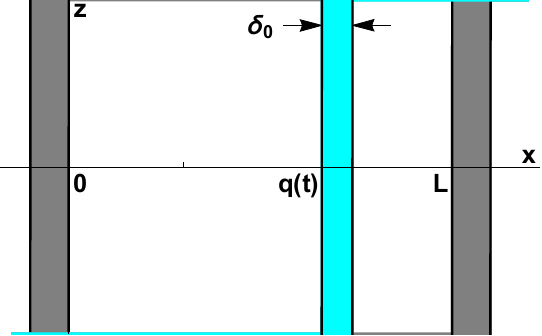}
\caption{\label{Figure1} Schematic of the MIM model. The interior of the cavity goes from $x=0$ to $x=L$ and the outer slabs correspond to the fixed, perfectly conducting mirrors. The mobile dielectric membrane is the slab inside the cavity. It has thickness $\delta_{0}$ and its midpoint is $q(t)$.}
\end{figure}

Consider a one-dimensional  cavity with two perfectly conducting, fixed  mirrors  and a mobile, dielectric membrane in between. We assume that the membrane is a linear, isotropic, non-magnetizable, non-conducting, and uncharged dielectric with thickness $\delta_{0}$ when it is at rest. In particular, the membrane is not restricted to small displacements near an equilibrium position. Figure~\ref{Figure1} gives a schematic representation of the system. The boundaries of the fixed mirrors are located at $x=0$ and $x=L$.  We denote by  $q(t)$ the midpoint of the membrane and refer to it as the position of the membrane. Moreover, $\chi (x)$ is the electric susceptibility of the membrane and  we assume that it is a nonnegative, piecewise continuous function with a piecewise continuous derivative such that $\chi[x-q(t)] = 0$ if $\vert x-q(t) \vert \geq \delta_{0}/2$. Also, the dielectric function  of the membrane is $\varepsilon:= 1 + 4 \pi \chi$ and we use Gaussian units. For a study of the coupled dynamics of the electromagnetic field inside the cavity and the moving membrane by means of self consistent, coupled equations see  \cite{PhysicaScripta}.

The electric and magnetic fields inside the cavity are derived from a vector potential
\begin{eqnarray}
\label{potenciales}
\mathbf{A}(x,t) \ = \ A_{0}(x,t)\mathbf{\hat{z}},
\end{eqnarray}
where $\mathbf{\hat{z}}$ is the unit vector in the direction of the positive $z$-axis. 

Let $\nu_{0}$ be a characteristic frequency of the electromagnetic field (there are no decay rates here), $\lambda_{0} = c/\nu_{0}$ with $c$ the speed of light in vacuum the corresponding characteristic wavelength, $A_{00}$ a characteristic value of $A_{0}(x,t)$, and $\nu_{\mbox{\tiny osc}}^{-1}$ the time-scale in which $q(t)$ varies appreciably. We measure  time in units of $\nu_{0}^{-1}$ and lengths in units of $\lambda_{0}$, so the dimensionless time $\tau$ and  position $\xi$ are 
\begin{eqnarray}
\label{XTadimensional}
\tau = \nu_{0}t , \, \qquad\xi = \frac{x}{\lambda_{0}} .
\end{eqnarray}
Also, we introduce following dimensionless quantities:
\begin{eqnarray}
\label{Adimensionales}
\xi_{L} &=& \frac{L}{\lambda_{0}}, \quad\quad\quad\quad \tilde{\chi}\left[ \xi -\tilde{q}(\tau) \right] = \chi \Big[ \lambda_{0}\left[ \xi -\tilde{q}(\tau) \right] \Big] , \quad\quad\quad \epsilon_{\mbox{\tiny pert}} = \frac{\nu_{\mbox{\tiny osc}}}{\nu_{0}} ,\cr
\tilde{\delta}_{0} &=& \frac{\delta_{0}}{\lambda_{0}} \ , \quad\quad\quad\quad \tilde{\epsilon}\left[ \xi -\tilde{q}(\tau) \right] = 1 + 4\pi\tilde{\chi}\left[ \xi -\tilde{q}(\tau) \right] , \cr
&& \cr
 \tilde{q}(\tau )& =& \frac{q(\nu_{0}^{-1}\tau )}{\lambda_{0}}, \quad\quad \tilde{A}_{0}(\xi , \tau ) = \frac{A_{0}(\lambda_{0}\xi , \nu_{0}^{-1}\tau )}{A_{00}} .
\end{eqnarray}
The quantities  $\xi_{L}$ and $\tilde{\delta}_{0}$ are the dimensionless length of the cavity and the dimensionless thickness of the membrane, while $\tilde{q}(\tau )$ is the dimensionless position of the membrane. Also, $\tilde{\chi}\left[ \xi -\tilde{q}(\tau) \right]$ and $\tilde{\epsilon}\left[ \xi -\tilde{q}(\tau) \right]$ are the electric susceptibility and the dielectric function of the membrane in terms of the  dimensionless argument $[ \xi -\tilde{q}(\tau) ]$, respectively, and $\tilde{A}_{0}(\xi , \tau )$ is the dimensionless potential.
The parameter $\epsilon_{\mbox{\tiny pert}}$ is the time-scale, $1/\nu_{0},$ in which the electromagnetic field evolves appreciably divided by the time-scale, $1/\nu_{\mbox{\tiny osc}},$ in which the membrane moves appreciably. In  the electromagnetic MIM model in \cite{Thompson}, taking the values from Table 1 one has $\nu_{0} = 5.63 \times 10^{14}$ Hz, $\nu_{\mbox{\tiny osc}} =   10^{5}$ Hz, and, consequently, $\epsilon_{\mbox{\tiny pert}} =1.8 \times 10^{-10}. $  In  the experiment \cite{MST}  in the microwave regime one has  $\nu_{0} = 5.153  $ GHz, $\nu_{\mbox{\tiny osc}} = 15.129 $ MHz, and, consequently, $\epsilon_{\mbox{\tiny pert}} =2.936  \times 10^{-3}.$

The general equation that governs the dynamics of $\tilde{A}_{0}(\xi,\tau)$ without any restriction on the  values of the velocity and the acceleration of the membrane was obtained in \cite{PhysicaScripta} by means of a  relativistic analysis.
 To first order in $\tilde{q}'(\tau)$ and $\tilde{q}''(\tau)$, the  equation is 
\begin{eqnarray}
\label{8}
\frac{\partial^{2}\tilde{A}_{0}}{\partial \xi^{2}}(\xi , \tau ) 
&=& \tilde{\epsilon}\left[ \xi - \tilde{q}(\tau) \right] \frac{\partial^{2}\tilde{A}_{0}}{\partial \tau^{2}}(\xi , \tau )  +8\pi \tilde{q}'(\tau )\tilde{\chi}\left[ \xi - \tilde{q}(\tau) \right] \frac{\partial^{2}\tilde{A}_{0}}{\partial \xi \partial \tau}(\xi , \tau ) \cr
&& +4\pi \tilde{q}''(\tau )\tilde{\chi}\left[ \xi - \tilde{q}(\tau) \right] \frac{\partial \tilde{A}_{0}}{\partial \xi }(\xi , \tau ) \ .
\end{eqnarray}
The boundary conditions that correspond to the perfectly conducting mirrors are
\begin{eqnarray}
\label{BCo}
\tilde{A}_{0}(0,\tau) \ = \ 0 \ , \ \ \tilde{A}_{0}(\xi_{L},\tau) \ = \ 0 \ .
\end{eqnarray}

\subsection{The angular frequencies and the modes of the cavity}\label{fixed}

We now present results from \cite{2016} on the evolution of the potential in the case where the membrane is fixed. 

Assume that the membrane is fixed at some point inside the cavity, 
\begin{eqnarray}
\label{Pi1}
\tilde{q}(\tau) = \tilde{q}_{0} \ \in \left( \frac{\tilde{\delta}_{0}}{2} , \ \xi_{L} - \frac{\tilde{\delta}_{0}}{2} \right) \ .
\end{eqnarray}
Since the velocity and the acceleration of the membrane are zero, (\ref{8}) reduces to the wave equation,
\begin{eqnarray}
\label{8f}
\frac{\partial^{2}\tilde{A}_{0}}{\partial \xi^{2}}(\xi , \tau ) 
&=& \tilde{\epsilon}\left( \xi - \tilde{q}_{0} \right) \frac{\partial^{2}\tilde{A}_{0}}{\partial \tau^{2}}(\xi , \tau ) \ .
\end{eqnarray}

We look for solutions of (\ref{8f}) that are periodic in time,
\beq\label{8f.a}
\tilde{A}_{0}(\xi , \tau )= e^{\pm i \omega(\tilde{q}_0) \tau} G(\xi,\tilde{q}_0).
\ene
The cavity has a countable set of dimensionless angular frequencies,  $\{ \omega_{n}(\tilde{q}_{0}) \}_{n=1}^{+\infty}$, and an associated set of cavity modes, $\{ G_{n}(\xi , \tilde{q}_{0}) \}_{n=1}^{+\infty}$, consisting of multiplicity one eigenfunctions  that are solutions to the following boundary value problem
\begin{eqnarray}
\label{BVPmodos}
\frac{\partial^{2}G_{n}}{\partial \xi^{2}}(\xi , \tilde{q}_{0}) &=& -\omega_{n}(\tilde{q}_{0})^{2} \tilde{\epsilon}( \xi -\tilde{q}_{0} ) G_{n}(\xi , \tilde{q}_{0}) \ , \cr
&& \cr
G_{n}(0, \tilde{q}_{0}) &=& 0 \ , \quad G_{n}(\xi_{L}, \tilde{q}_{0}) = 0 \ .
\end{eqnarray}
We  order the angular frequencies  in such a way  that $\omega_{m}(\tilde{q}_{0}) < \omega_{n}(\tilde{q}_{0})$ if $m < n$.
Moreover, $\omega_{n}(\tilde{q}_{0}) >0$ and $\lim_{n\to\infty} \omega_{n}(\tilde{q}_{0})= \infty.$ In addition, there are no  crossings: if $n \neq m$, then  $\omega_{n}(\tilde{q}_{0})\neq \omega_{m}(\tilde{q}_{0})$ for all $ \tilde{q}_{0} \ \in \left( \frac{\tilde{\delta}_{0}}{2} , \ \xi_{L} - \frac{\tilde{\delta}_{0}}{2} \right).$ 
The set of cavity modes is an orthonormal basis of real-valued functions in the Hilbert space $ L^2_{\tilde{q}_0}$ of all complex-valued, Lebesgue square-integrable functions on $(0,\xi_L)$ with the scalar product
$$
\left( f,g  \right)_{\tilde{\varepsilon}}:= \int_{0}^{\xi_{L}}d\xi \ \tilde{\epsilon}(\xi - \tilde{q}_{0})\overline{f (\xi , \tilde{q}_{0})} g(\xi , \tilde{q}_{0}). 
$$ 

\subsection{The multiple scales approximation}\label{multiple}

Following \cite{2016}, we consider now  the situation where the membrane is allowed to move. 

For each fixed $\tau$ we denote by 
$\{ \omega_{n}[\tilde{q}(\tau) ] \}_{n=1}^{+\infty}$ \ the instantaneous angular frequencies of the cavity and by    $\{ G_{n}[ \xi , \tilde{q}(\tau) ] \}_{n=1}^{+\infty}$  the associated  set of instantaneous modes. For simplicity we refer to them as the cavity angular frequencies and modes. Using that for each fixed $\tau$ the modes of the cavity are an orthonormal basis, we expand $\tilde{A}_{0}(\xi , \tau) $ as
\begin{eqnarray}
\label{ExpansionEnModos}
\tilde{A}_{0}(\xi , \tau) &=& \sum_{n=1}^{+\infty} c_{n}(\tau ) G_{n}[ \xi , \tilde{q}(\tau) ] .
\end{eqnarray}
Introducing (\ref{ExpansionEnModos}) into  (\ref{8})  we deduce the equations for the coefficients $c_{n}(\tau )$:
\begin{eqnarray}
\label{10}
&& c_{m}''(\tau) + \omega_{m}[ \tilde{q}(\tau) ]^{2} c_{m}(\tau )= \cr
&& - \sum_{n=1}^{+\infty} \Gamma_{mn}[ \tilde{q}(\tau) ] \Big[ \ 2\tilde{q}'(\tau )c_{n}'(\tau) + \tilde{q}''(\tau )c_{n}(\tau) \ \Big] \ .
\end{eqnarray}
Here we introduced the following  quantities:
\begin{eqnarray}
\label{ModeDependent}
\Omega_{mn}\left[ \tilde{q}(\tau ) \right] &=& \int_{0}^{\xi_{L}} d\xi \ \tilde{\epsilon}\left[ \xi - \tilde{q}(\tau) \right] G_{m}\left[ \xi , \tilde{q}(\tau) \right] \frac{\partial G_{n}}{\partial \tilde{q}(\tau )}\left[ \xi , \tilde{q}(\tau) \right] \ , \cr
&& \cr
\theta_{mn}\left[ \tilde{q}(\tau ) \right] &=& \int_{0}^{\xi_{L}} d\xi \ 4 \pi \tilde{\chi}\left[ \xi - \tilde{q}(\tau) \right] G_{m}\left[ \xi , \tilde{q}(\tau) \right] \frac{\partial G_{n}}{\partial \xi }\left[ \xi , \tilde{q}(\tau) \right] \ , \cr
&& \cr
\Gamma_{mn}\left[ \tilde{q}(\tau ) \right] &=& \Omega_{mn}\left[ \tilde{q}(\tau ) \right] + \theta_{mn}\left[ \tilde{q}(\tau ) \right] \ .
\end{eqnarray}
The system of ordinary differential equations (\ref{10}) for the expansion coefficients $c_n(\tau)$ in (\ref{ExpansionEnModos}) is equivalent to the wave equation (\ref{8}). As it is the case with (\ref{8}),  the system (\ref{10}) is accurate to first order in $\tilde{q}'(\tau)$ and  $\tilde{q}''(\tau)$. 

We make the following four assumptions \cite{2016}:
 \begin{enumerate}
 \item The functions $ \tilde{q}'(\tau)$ and  $ \tilde{q}''(\tau)$ are small. These conditions are necessary so that 
 the system (\ref{10}) is correct to first order in  $ \tilde{q}'(\tau)$ and  $ \tilde{q}''(\tau).$ 
\item $0 < \epsilon_{\mbox{\tiny pert}} \ll 1$. 
This assumption  implies  that there are two clearly defined and separate time-scales: a fast time-scale $1/\nu_{0}$ in which the electromagnetic field evolves and a slow time-scale $1/\nu_{\mbox{\tiny osc}}$ in which the membrane moves. 
\item The initial conditions for the coefficients $c_{m}(\tau)$ of the modes are
\begin{eqnarray} 
\label{13}
c_{m}(0) &=& g_{0N}\delta_{mN} \ , \quad c_{m}'(0) = g_{1N}\delta_{mN} \ ,
\end{eqnarray}
where $g_{0N}$ and $g_{1N}$ are real numbers and $N$ is a fixed, positive integer.
 These  initial conditions  indicate that only mode $N$ is initially excited. 
\item The membrane is initially at rest, i.e., $\tilde{q}'(0) = 0$.
\end{enumerate}
These  assumptions correspond to the following physical situation: the membrane starts to move from rest and the field is initially found in one of the modes of the cavity. 

Let
\begin{eqnarray}
\label{EscalaRapida}
t_{1N}(\tau) &=& \int_{0}^{\tau} \omega_{N}\left[ \tilde{q}(\rho) \right] d\rho \ , \cr
\alpha_{N} \left[ \tilde{q}(\tau) \right] &=& \sqrt{ \frac{\omega_{N}\left[ \tilde{q}(0) \right]}{\omega_{N}\left[ \tilde{q}(\tau) \right]} } \ \mbox{exp}\left\{ - \int_{\tilde{q}(0)}^{\tilde{q}(\tau )} dy \ \Gamma_{NN}(y) \right\} \ , \cr
&& \cr
b_{N0} &=& \left\vert \frac{g_{0N}}{2} + \frac{i g_{1N}}{2\omega_{N}\left[ \tilde{q}(0) \right]}  \right\vert \ , \cr
&& \cr
\Theta_{N0} &=& \mbox{arg}\left[ \frac{g_{0N}}{2} + \frac{i g_{1N}}{2\omega_{N}\left[ \tilde{q}(0) \right]} \right] \ .
\end{eqnarray} \ 
Then, the multiple scales solution is given by \cite{2016}
\begin{eqnarray}
\label{A02}
\tilde{A}_{0}^{(2)}(\xi , \tau) &=&  \alpha_{N} \left[ \tilde{q}(\tau) \right] b_{N0} e^{-i\left[ t_{1N}(\tau) - \Theta_{N0} \right] } G_{N}[ \xi , \tilde{q}(\tau) ] \times \cr
&& \quad\quad \left\{1+ i\tilde{q}'(\tau) \frac{\omega_{N}'\left[ \tilde{q}(\tau) \right]}{4 \omega_{N}\left[ \tilde{q}(\tau) \right]^{2}} \right\}  \cr
&& +  \alpha_{N} \left[ \tilde{q}(\tau) \right] b_{N0} e^{-i\left[ t_{1N}(\tau) - \Theta_{N0} - \pi/2 \right] } \tilde{q}'(\tau) \times \cr
&&  \sum_{m=1 \atop m\not=N}^{+\infty}  \frac{2 \Gamma_{mN}\left[ \tilde{q}(\tau) \right] \omega_{N} \left[ \tilde{q}(\tau) \right] }{\omega_{m}\left[ \tilde{q}(\tau ) \right]^{2} - \omega_{N}\left[ \tilde{q}(\tau ) \right]^{2}}  G_{m}[ \xi , \tilde{q}(\tau) ] \cr
&& + c.c.
\end{eqnarray}
Observe that an electromagnetic field initially in mode $N$ will follow mode $N$, provided that  the membrane moves slowly. In addition, only modes \ $n\not= N$ \ with $\omega_{n}[\tilde{q}(\tau)]$ in a small band around  $\omega_{N}[\tilde{q}(\tau)]$ can have a non-negligible excitation. 

 \section{The case of a slab membrane}
 \label{slab}
In this section we present the new results. We consider the particular case where the electric susceptibility is 
 \begin{eqnarray}
\label{Susceptibilidad}
\tilde{\chi} ( \xi - \tilde{q}_0 ) &=&
\left\{
\begin{array}{cc}
\chi_{0} & \mbox{if} \ \ \vert \xi - \tilde{q}_0 \vert \ < \ \frac{\tilde{\delta}_{0}}{2} \ , \cr
0 & \mbox{elsewhere.}
\end{array}
\right.
\end{eqnarray}
We assume that the membrane is fixed  at $\tilde{q}_0$ as in (\ref{Pi1}) and we introduce the following convenient notation
\beq\label{s.1}
\alpha:=\sqrt{1+ 4\pi \chi_0}, \quad 
\beta:= \frac{\tilde{q}_0- \tilde{\delta}_0/2}{\xi_L}.
\ene
Hence,
\beq\label{s.3}
\tilde{q}_0- \tilde{\delta}_0/2 = \beta \xi_L, \qquad 0 < \beta < 1- \frac{\tilde{\delta}_0}{\xi_L}.
\ene
Imposing that the Dirichlet boundary condition is satisfied at $ \xi=0$ and $\xi=\xi_L$ and  the continuity of $G_{n}(\xi ,\tilde{q}_{0})$ and its first partial derivative with respect to $\xi$ at the boundaries of the membrane $\xi = \tilde{q}_{0} \pm \tilde{\delta}_{0}/2$, we  calculate explicitly the modes in terms of elementary functions and we obtain an implicit equation for the calculation of the angular frequencies of the cavity.  Namely,
\beq\label{s.4}
G_{n}(\xi , \tilde{q}_{0}) = \frac{1}{\sqrt{\mathcal N_n(\tilde{q}_0)}} \widehat{G}_{n}(\xi , \tilde{q}_{0}),
\ene
with
\beq
\label{uno}
\widehat{G}_{n}(\xi , \tilde{q}_{0}) =  \sin[\omega_n(\tilde{q}_0) \xi] \quad \mathrm{if} \quad 0 \leq \xi \leq \tilde{q}_0- 
\frac{\tilde{\delta}_0}{2}, 
\ene
and
\begin{eqnarray}
\label{dos}
\widehat{G}_{n}(\xi , \tilde{q}_{0}) &=& \cos[\omega_n(\tilde{q}_0) \alpha(\xi-\xi_L\beta)] \sin[\xi_L\omega_n(\tilde{q}_0)\beta] \cr
&& + \frac{1}{\alpha} \sin[\omega_n(\tilde{q}_0) \alpha(\xi-\xi_L\beta)] \cos[\xi_L\omega_n(\tilde{q}_0) \beta] \cr
&\mathrm{if}&\left|\xi- \tilde{q}_0 \right|\leq \tilde{\delta}_0/2,
\end{eqnarray}
and
\begin{eqnarray}
\label{s.5}
&& \widehat{G}_{n}(\xi , \tilde{q}_{0}) \cr
&=& \frac{1}{4\alpha}\Bigg\{ (1+\alpha)^2  \sin\left\{ \omega_n(\tilde{q}_0)\left[ \xi+\tilde{\delta}_0 (\alpha-1) \right] \right\}  \cr
&& -  (\alpha-1)^2\sin\left\{  \omega_n(\tilde{q}_0)\left[ \xi-\tilde{\delta}_0 (\alpha+1) \right] \right\} \cr
&& - (\alpha^2-1) \sin\left\{ \omega_n(\tilde{q}_0) \left[ \xi+\tilde{\delta}_0(\alpha-1)- 2\xi_L \beta \right] \right\}  \cr
&& + (\alpha^2-1) \sin\left\{ \omega_n(\tilde{q}_0) \left[ \xi-\tilde{\delta}_0(\alpha+1)- 2\xi_L \beta\right] \right\} \Bigg\} \cr
&\mathrm{if}&\tilde{q}_0+ \tilde{\delta}_0/2 \leq \xi \leq \xi_{L} .
\end{eqnarray}
The coefficient $\mathcal N_n(\tilde{q}_0)$ is obtained by requiring that ${G}_{n}(\xi , \tilde{q}_{0})$ be normalized to one. It is given by
\beq\label{s.6}
\mathcal N_n(\tilde{q}_0):= \int_0^{\xi_L}\, \tilde{\varepsilon}(\xi-\tilde{q}_0) |\widehat{G}_{n}(\xi , \tilde{q}_{0})|^2 d\xi ,
\ene
and it can be evaluated explicitly using (\ref{s.6}) and a symbolic programming language like Mathematica. The expression is, however, rather complicated and we decided not to include it. Instead, in the numerical computations we evaluate it whenever it is needed. Moreover, we have the following implicit equation to evaluate the angular frequencies of the cavity:
\beq\label{s.7}
\begin{array}{l}
  - (\alpha-1)^2\sin\left\{ \omega(\tilde{q}_0)\left[ \xi_L-\tilde{\delta}_0(\alpha+1) \right] \right\} \\
- 2(\alpha^2-1)  \cos\left[ \omega(\tilde{q}_0)\left( \xi_L- \tilde{\delta}_0 -2\xi_ L\beta \right) \right] \sin\left[ \omega(\tilde{q}_0)  \tilde{\delta}_0 \alpha \right] \\
+ (\alpha+1)^2 \sin\left\{ \omega(\tilde{q}_0)\left[ \xi_L+\tilde{\delta}_0(\alpha-1) \right] \right\} =0.
\end{array}
\ene 
Recall that the angular frequencies of the cavity are positive, so we have to look for positive solutions to (\ref{s.7}). Below we find an explicit solution to (\ref{s.7}). For further explicit solutions see  the next section. 

Assume that the membrane is in the middle of the cavity. Then,
$$
\tilde{q}_0= \frac{\xi_L}{2} = \xi_L\beta+\frac{\tilde{\delta}_0}{2}, 
$$
and
\beq\label{s.7a}
2 \xi_L \beta= \xi_L- \tilde{\delta}_0.
\ene
Using (\ref{s.7a}), it follows from a simple computation that (\ref{s.7}) can be written as,
\beq\label{s.7b}\begin{array}{l}
(\alpha^2+1) \cos[\omega(\tilde{q}_0)(\xi_L- \tilde{\delta}_0)] \sin[\omega(\tilde{q}_0)  \tilde{\delta}_0\alpha ]- (\alpha^2-1)  \sin[\omega(\tilde{q}_0)  \tilde{\delta}_0\alpha ]+ \\\\ 2 \alpha \sin[\omega(\tilde{q}_0)(\xi_L- \tilde{\delta}_0)] \cos[\omega(\tilde{q}_0)  \tilde{\delta}_0\alpha ]=0.
\end{array}
 \ene
 Take
 \beq\label{s.7c}
 \omega(\tilde{q}_0)= \omega_n(\tilde{q}_0):= \frac{1}{\tilde{\delta}_0\alpha } (2n+1) \frac{\pi}{2}, \qquad n=0,1,\dots.
 \ene
 Then, (\ref{s.7b}) reduces to,
 \beq\label{s.7d}
  \cos[\omega_n(\tilde{q}_0)(\xi_L- \tilde{\delta}_0)] = \frac{\alpha^2-1}{\alpha^2+1}= \frac{4\pi \chi_0}{2+4\pi \chi_0}.
 \ene
 Denote,
 $$
 \gamma:= \arccos\left[ \frac{4\pi \chi_0}{2+4\pi \chi_0}\right].
 $$
 Then, by (\ref{s.7c}), (\ref{s.7d}) is equivalent to,  
 \beq\label{s.7e}
  \tilde{\delta}_0=  \tilde{\delta}_{0,n} := \xi_L \frac{(2n+1)\pi}{(2n+1)\pi+ 2 \alpha \gamma},\qquad n=0,1,\dots.
  \ene
  Further, introducing (\ref{s.7e}) into (\ref{s.7c}) we get,
  \beq\label{s.7f}  \omega(\tilde{q}_0)= \omega_n(\tilde{q}_0)=\frac{1}{\xi_L}\left[ \gamma+ \frac{1}{\alpha} (2n+1)\frac{\pi}{2}\right], \quad n=0,1,\dots.
  \ene
  Hence, we have proven that, for $ \tilde{q}_0= \xi_L/2,$ the angular frequencies (\ref{s.7f}) are explicit solutions to 
  (\ref{s.7}) for the widths of the membrane given by (\ref{s.7e}).

 \subsection{Structural angular frequencies}\label{structural}
In this subsection we give explicitly a sequence of {\it structural angular frequencies} that have the remarkable property that they are independent of the position of the membrane inside the cavity. More precisely, for each fixed value of the length of the cavity and each fixed value of the electric susceptibility of the membrane, we find a sequence of widths  of the membrane such that, for each width in the sequence, there is a cavity angular frequency that is independent of the position of the membrane inside the cavity, i.e., such that it is a cavity angular  frequency for all positions of the membrane inside the cavity. Furthermore, these cavity angular frequencies are the only ones that have the remarkable property of being independent of the position of the membrane inside the cavity. 
\\
\\
\noindent{\bf Theorem}
Let us consider a fixed $ \xi_L >0$ and a fixed $ \chi_0 > 0.$ For $k=1,2, \dots$ and $n = 1,2, \dots$ we define
\beq\label{s.8}
\omega_{n,k}:= \frac{(n \alpha+k)\pi }{\xi_L \alpha},  
\ene
and
\beq\label{s.9}
\tilde{\delta}_{0,n,k} :=\xi_ L \frac{k}{n\alpha+k}  < \xi_L .
\ene
Then, $\omega_{n,k}$  is a solution to (\ref{s.7}), i.e., it is one of the cavity's angular frequencies with $\tilde{\delta}_0=\tilde{\delta}_{0,n,k} $ as in (\ref{s.9}) with the same $n,k $ for all $ \tilde{\delta}_0/2 < \tilde{q}_0 < \xi_L-\tilde{\delta}_0/2$. Moreover, the only cavity angular frequencies that satisfy (\ref{s.7}) with a fixed $\tilde{\delta}_0 \in (0 ,\xi_L)$ for all $\tilde{q}_0$ in a nontrivial interval contained in $(\tilde{\delta}_0/2, \xi_L- \tilde{\delta}_0/2)$ are the ones given by (\ref{s.8}) with $\tilde{\delta}_0= \tilde{\delta}_{0,n,k}$ given by (\ref{s.9}) with the same
$n,k.$
\\
\\
\noindent{\it Proof:}
Let us take in (\ref{s.7}) 
 \beq\label{s.10}
 \omega=\frac{ k \pi}{\alpha \tilde{\delta}_{0}}, \qquad k= 1,2,\dots.
 \ene
 Then, $\sin(\omega \tilde{\delta}_{0} \alpha)=0$ and  (\ref{s.7}) simplifies to
\beq\label{s.11}
\begin{array}{l}
 (\alpha+1)^2 \sin\left\{ \omega\left[ \xi_L+\tilde{\delta}_{0}(\alpha-1) \right] \right\} \\\\
 = (\alpha-1)^2 \sin\left\{ \omega\left[ \xi_L-\tilde{\delta}_{0}(\alpha+1) \right] \right\} .
\end{array}
\ene 
Further, introducing (\ref{s.10}) into (\ref{s.11}) we get
\beq\label{s.12}
4 \alpha\sin\left( \omega \xi_L- k\frac{\pi}{\alpha} \right) = 0.
\ene
Hence, for $k,n = 1,2, \dots$ one has
\beq\label{s.13}
w= \omega_{n,k}:= \frac{(n \alpha+k)\pi }{\xi_L \alpha} .
\ene
Finally, using  (\ref{s.10}) and (\ref{s.13}) we obtain (\ref{s.9}). Note that the case  $n=0$ in (\ref{s.8}), (\ref{s.9}) is excluded because we need that $ \tilde{\delta}_0 <\xi_ L.$  

Now, assume that $w>0$ satisfies (\ref{s.7}) for a fixed $\xi_L$ and that (\ref{s.7}) holds for all  $\tilde{q}_0 \in (a,b)\subset (\tilde{\delta}_0/2, \xi_L- \tilde{\delta}_0/2)$ for some $a<b$. Then, we can take the derivative of both sides of (\ref{s.7}) with respect to 
$\beta $ to obtain that $\sin( \omega \tilde{\delta}_{0} \alpha )=0$. Hence, (\ref{s.10}) holds and, as above, we prove (\ref{s.8}) and (\ref{s.9}). 

\bull

Observe that the widths given in (\ref{s.9}) are not all different from each other. In fact, fixing the value of the width in the left-hand side of (\ref{s.9}) gives a relation between $n$ and $k.$

Note that keeping $k$ fixed in (\ref{s.9}) and taking $n$ large we can make the membrane as thin as we wish, but  by (\ref{s.8}) the corresponding {\it structural angular frequency} becomes very large.

We proceed to prove that the {\it structural angular frequencies} are always larger than the fundamental angular frequency
$\omega_1(\tilde{q}_0)$ with $\tilde{q_0} \in  \left( \frac{\tilde{\delta}_{0}}{2} , \ \xi_{L} - \frac{\tilde{\delta}_{0}}{2} \right)$ and that  they are separated from the fundamental frequency by as least $\pi/(\xi_L\alpha)$. 
This is a simple consequence of the min-max principle in the theory of selfadjoint operators in Hilbert space. We denote by $L^2(0,\xi_L)$ the standard space of Lebesgue square-integrable functions on $(0,\xi_L).$   As mentioned in Section~ \ref{fixed},  we denote by $L^2_{\tilde{q}_0}$ with 
$ \tilde{q}_0 \in  \left( \frac{\tilde{\delta}_{0}}{2} , \ \xi_{L} - \frac{\tilde{\delta}_{0}}{2} \right)$ the space $L^2(0, \xi_L)$ endowed with the scalar product 
 $$
 (\phi,\psi)_{L^2_{\tilde{q}_0}}:=\int_{0}^{\xi_{L}}d\xi \ \tilde{\epsilon}(\xi - \tilde{q}_{0}) \overline{\phi(\xi)} \psi(\xi).
 $$
By $H_2$ we designate the second Sobolev space on  $(0,\xi_L)$ and by $H_{1,0}$ the first Sobolev space of functions on  $(0,\xi_L)$  that vanish for $\xi=0$ and $\xi= \xi_L.$ For the definition and properties of these spaces see \cite{Sobolev}. We define the following selfadjoint, positive operator $\mathbb A_{\tilde{q}_0}$ in $L^2_{\tilde{q}_0},$
$$   
\mathbb A_{\tilde{q}_0}\phi:= - \frac{1}{\tilde{\varepsilon}(\xi-\tilde{q}_0)} \frac{d^2}{d\xi^2} \phi,
$$
with the following domain,
$$
D\left(\mathbb A_{\tilde{q}_0} \right):= H_2 \cap H_{1,0}.
$$
As proved in \cite{2016} the spectrum of $\mathbb A_{\tilde{q}_0}$ consists of eigenvalues of multiplicity one that coincide with the square of the angular frequencies. Moreover, the quadratic form of 
$\mathbb A_{\tilde{q}_0}$ is given by
$$
h_{\tilde{q}_0}(\phi, \psi):= \int_0^{\xi_L}d \xi \overline{\phi'(\xi)} \psi'(\xi), 
$$
with domain
$$
D(h_{\tilde{q}_0}):= H_{1,0}.
$$
Let us denote by $\mathbb B$ the Dirichlet Laplacian in  $L^2(0,\xi_L),$
$$
\mathbb B\phi:= - \frac{d^2}{d\xi^2} \phi,
$$
with  domain
$$
D\left(\mathbb{B} \right):= H_2 \cap H_{1,0}.
$$
$\mathbb{B}$ is selfadjoint, positive, and its spectrum, $\{ (\frac{k \pi}{\xi_L})^2 \}_{k=1}^\infty$, consists of the eigenvalues of multiplicity one. Further, the quadratic form of $\mathbb{B}$ is given by,
$$
h_\mathbb{B}(\phi, \psi):= \int_0^{\xi_L}d \xi \overline{\phi'(\xi)} \psi'(\xi), 
$$
with domain
$$
h_\mathbb{B}:= H_{1,0}.
$$

Note that,
$$
h_{\tilde{q}_0}(\phi, \psi)= h_\mathbb{B}(\phi, \psi), \qquad \phi, \psi \in H_{1,0},
$$
but that $\mathbb A_{\tilde{q}_0}$ and $\mathbb B$ act in different Hilbert spaces, namely, in $L^2_{\tilde{q}_0}$ and $L^2(0,\xi_L)$, respectively.
By the min-max principle (see Theorem XIII.2 in page 78 of \cite{rs})
\beq\label{s.14}
\begin{array}{l}
\omega_1(\tilde{q}_0)^2= \inf_{\phi \in H_{1,0}, \phi  \neq 0}  \frac{\left( \phi',\phi'\right)_{L^2(0,\xi_L)}}{ \|\phi\|^2_{L^2_{\tilde{q}_0} }}\\
\leq   \inf_{\phi \in H_{1,0}, \phi  \neq 0}  \frac{\left( \phi',\phi'\right)_{L^2(0,\xi_L)}}{ \|\phi\|^2_{L^2(0,\xi_L)}}=  (\frac{ \pi}{\xi_L})^2,
\end{array}
\ene
where we use that 
$$
 \|\phi\|_{L^2_{\tilde{q}_0}} \geq  \|\phi\|_{L^2(0,\xi_L)},
 $$
 and that $  (\frac{ \pi}{\xi_L})^2$ is the smallest eigenvalue of $\mathbb{B}$. Then, by (\ref{s.8}) and (\ref{s.14}), for $ k,n= 1,2,\dots,$
 \beq\label{s.15}\begin{array}{l}
 \omega_{n,k} \geq \omega_{1,1} = \frac{\pi}{\xi_L}+ \frac{\pi}{\xi_L\alpha} \geq \omega_1(\tilde{q}_0)+\frac{\pi}{\xi_L\alpha} , \\
 \tilde{q}_0 \in  \left( \frac{\tilde{\delta}_{0}}{2} , \ \xi_{L} - \frac{\tilde{\delta}_{0}}{2} \right).
 \end{array}
 \ene
 Hence, the {\it structural angular frequencies} are strictly larger that the fundamental angular frequency
$ \omega_1(\tilde{q}_0)$ and they are separated from the fundamental frequency by at least $\frac{\pi}{\xi_L\alpha}.$ 

\begin{figure}[H]
\includegraphics[width=8cm]{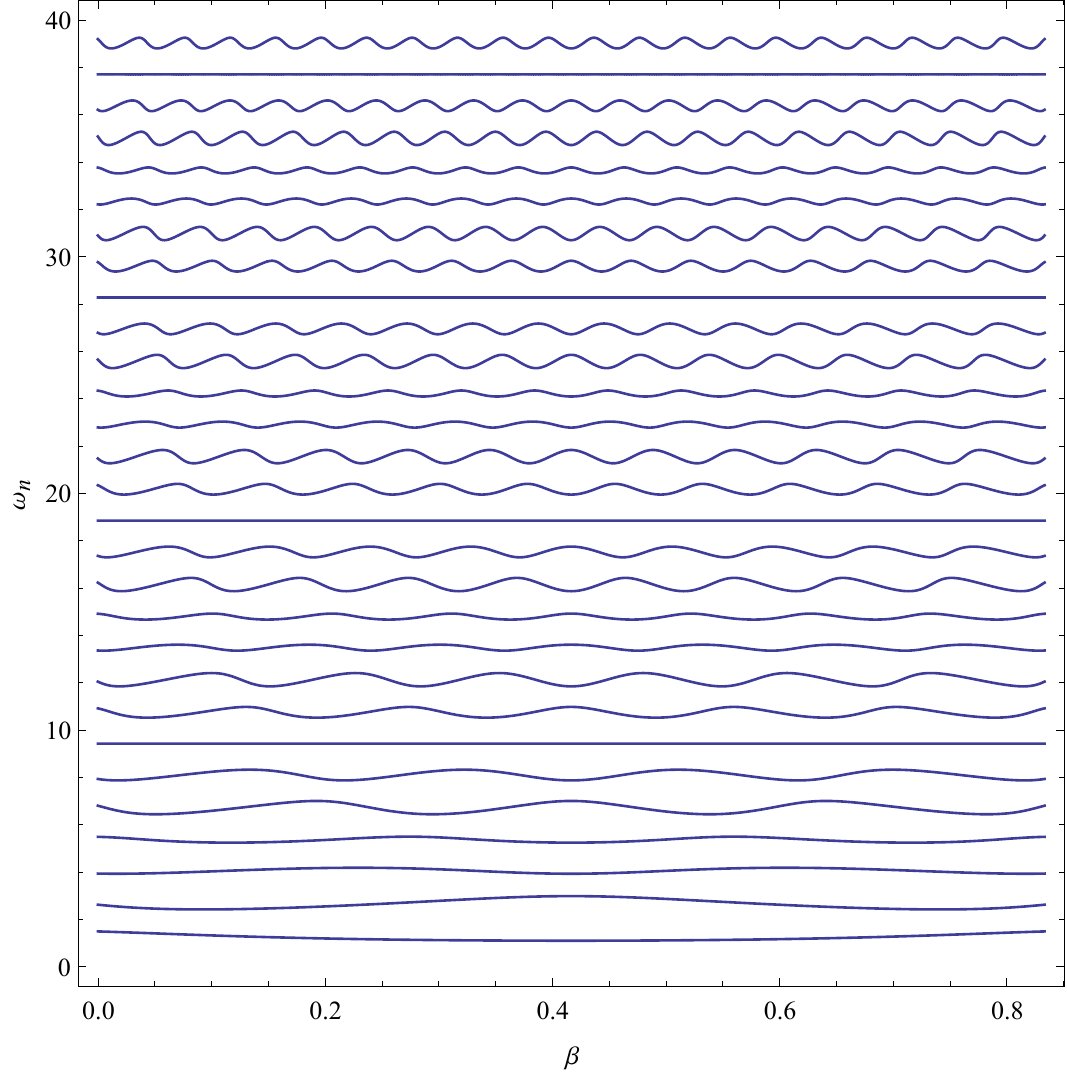}
\caption{\label{Figure2} Numerical computation of the lowest $29$ angular frequencies using the implicit equation (\ref{s.7}), as a function of the  parameter $\beta:= \frac{\tilde{q}_0- \tilde{\delta}_0/2}{\xi_L}$. The width of the membrane is $\tilde{\delta}_0,$ the  position of the  left side of the  membrane is $\tilde{q}_0- \frac{1}{2} \tilde{\delta}_0,$ and the length  of  the cavity is $\xi_L.$ Note that, $ 0 <  \beta < 1- \frac{\tilde{\delta}_0}{\xi_L}.$ We have taken $ \xi_L=2, \tilde{\delta}_0= \frac{1}{3},$ and $\alpha:=\sqrt{1+4\pi\chi_0}=2.$ The first four structural angular frequencies appear as straight  horizontal lines.  }
\end{figure}

In Figure~\ref{Figure2} we show the lowest $29$ angular frequencies of the cavity computed numerically using the implicit equation (\ref{s.7}) as a function of the  parameter $\beta:= \frac{\tilde{q}_0- \tilde{\delta}_0/2}{\xi_L}.$  We take $\tilde{\delta}_0= 1/3, \alpha=2,$ and $\xi_L=2.$ For these values (\ref{s.8}) gives the sequence of {\it structural angular frequencies}
$\omega_{5l,2l}= 3 \pi l, l=1,\dots$ . The lowest four {\it structural angular frequencies} appear as horizontal straight lines in Figure~\ref{Figure2} .

\subsection{The case of a thin membrane}

In this subsection we consider approximate solutions when the width $ \tilde{\delta}_0$ of  the membrane is small.
This case is important, as often in the experiments and in the applications the width of the membrane is is very small. For example, in \cite{Thompson} the width of the membrane in units of the cavity's length is $ \tilde{\delta}_0= 0.74 \times10^{-6}.$ Below we denote by $G_{n,l}(\xi , \tilde{q}_{0})$ and $ \widehat{G}_{n,l}(\xi , \tilde{q}_{0}),$ respectively, the approximate normalized solution and the approximate unnormalized solution to first order in $\tilde{\delta}_{0}$. Further we denote by  ${\mathcal N_{n,l}(\tilde{q}_0)}$ the norming coefficient of the approximate normalized solution $G_{n,l}(\xi , \tilde{q}_{0})$.
Approximating (\ref{s.5}) to first order in  $ \tilde{\delta}_0$ (note that (\ref{uno}) and (\ref{dos}) do not depend explicitly on $\tilde{\delta}_0$),  we obtain 

\beq\label{s.16}
G_{n,l}(\xi , \tilde{q}_{0}) = \frac{1}{\sqrt{\mathcal N_{n,l}(\tilde{q}_0)}} \widehat{G}_{n,l}(\xi , \tilde{q}_{0}),
\ene
where $\widehat{G}_{n,l}(\xi , \tilde{q}_{0})$ is exactly equal to $\widehat{G}_{n}(\xi , \tilde{q}_{0})$ in (\ref{uno}) and (\ref{dos}) for $\xi \leq \tilde{q}_{0} + \tilde{\delta}_{0}/2$ and
\beq
\begin{array}{l}\label{s.17}
\widehat{G}_{n,l}(\xi , \tilde{q}_{0})= \sin[\omega_n(\tilde{q}_0) \xi]+ \frac{1}{2} \tilde{\delta}_0\Big\{ \omega_n(\tilde{q}_0)(-1+\alpha^2)  \\\\
 \cos[\omega_n(\tilde{q}_0) \xi]+ (1-\alpha^2) \omega_n(\tilde{q}_0)
\cos\left[ \omega_n(\tilde{q}_0) (\xi- 2 \xi_L \beta) \right] \Big\}, \\\\  \mathrm{for} \quad \tilde{q}_0+ 
\frac{\tilde{\delta}_0}{2}\leq   \xi \leq  \xi_L.
\end{array}
\ene
By the Taylor expansion, the error made in approximating  (\ref{s.5}) by (\ref{s.17}) is smaller than 
$\frac{1}{2} \alpha  (\alpha^2-1) [\omega_n(\tilde{q}_0)\, \tilde{\delta}_0 ]^2.$
The norming coefficient  $\mathcal N_{n,l}$ is defined as
\beq\label{s.17b}
\mathcal N_{n,l}(\tilde{q}_0):= \int_0^{\xi_L}\, \tilde{\varepsilon}(\xi-\tilde{q}_0) |\widehat{G}_{n,l}(\xi , \tilde{q}_{0})|^2 d\xi,
\ene

\beq\label{s.18}
\begin{array}{l}
\displaystyle \sqrt{\mathcal N_{n,l}(\tilde{q}_0)}=\frac{1}{2} \Bigg\{
2 \xi_L  -\frac{\sin[2 \xi_L \omega_n(\tilde{q}_0)]}{ \omega_n(\tilde{q}_0)}
+ \tilde{\delta}_0 (-1+\alpha^2)\\\\\Big(1- \cos[2\xi_L\omega_n(\tilde{q}_0) ] + 
 \cos[2\xi_L\omega_n(\tilde{q}_0) (-1+\beta)]- \\\\ \cos[2\xi_L\omega_n(\tilde{q}_0) \beta]  + 2 \xi_L \omega_n(\tilde{q}_0) (-1+\beta) \sin[2 \xi_L\omega_n(\tilde{q}_0) \beta] \Big) \Bigg\}^{1/2}.
 \end{array}
\ene

 Moreover, to first order in $  \tilde{\delta}_0$  equation (\ref{s.7}) takes the form 
\beq\label{s.19}\begin{array}{l}
\sin[\xi_L\omega(\tilde{q}_0) ] \\\\ + \omega(\tilde{q}_0) \tilde{\delta}_0 (-1+\alpha^2) \sin[\xi_L \omega(\tilde{q}_0)(-1+\beta)] \sin[\xi_L\omega(\tilde{q}_0) \beta]=0.
\end{array}
 \ene
By the Taylor expansion,  the error made in approximating (\ref{s.7}) by (\ref{s.19}) is smaller than  $ 2 \alpha (\alpha^2-1) [ \omega(\tilde{q}_0) \tilde{\delta}_0]^2.$  Hence, for this error to be small for the angular frequency $\omega_n(\tilde{q}_0)$ one needs 
\beq\label{s.23b}
\Delta_n:= 2 \alpha  (\alpha^2-1) \left[ \omega_n(\tilde{q}_0)\, \tilde{\delta}_0 \right]^2 << 1, \quad n=1,\dots.
\ene
Note that  (\ref{s.19}) has the following explicit solutions
\beq\label{s.20}
\omega_{n}(\tilde{q}_0)=   \frac{n\pi}{\xi_L },
\ene
where we choose  $\beta$ as follows
\beq\label{s.20b}
\beta= \beta_{n,k}:= 1-\frac{k}{n},  \quad \frac{\tilde{\delta}_0}{\xi_L} < \frac{k}{n}<1, \quad n,k =1,2\dots.
\ene
This means that for each value of the quotient  $\tilde{\delta}_0/\xi_L,$ if  we  pick the parameter $\beta$ as in (\ref{s.20b}) the angular frequency is given by (\ref{s.20}). Recall that according to (\ref{s.3}) $\beta$ fixes the position of the left side of the membrane.

 We now proceed to obtain approximate angular frequencies.
When $\tilde{\delta}_0=0 ,$ there is no membrane and the exact  angular frequencies of the cavity are
$ \omega_n= n \pi/\xi_L$ with $n=1,2,\dots.$  As we are considering thin membranes, it is natural to look for approximate
solutions to  (\ref{s.19}) of the form
\beq\label{s.21}
\omega_{n,a}(\tilde{q}_0) = \frac{n \pi}{\xi_L}+ d_n,
\ene
\beq\label{s.21b}
|d_n|  << 1,\qquad n=1,\dots .
\ene
Introducing (\ref{s.21}) into (\ref{s.19}), keeping only the terms of order zero and one in $d_n$
and solving for $d_n$ we obtain,
\beq\label{s.22}
\begin{array}{l}
d_n= - \frac{1}{2\xi_L^2} n  \tilde{\delta}_0  \pi (\alpha^2-1) \left[1-\cos(2n \pi \beta) \right] ,\quad\quad  n=1,\dots.
\end{array}
\ene
Since (\ref{s.19}) is an approximation to first order in  $\tilde{\delta}_0$ of the exact implicit equation (\ref{s.7}),  by consistency, in (\ref{s.22}) we have kept only the terms of first order in $\tilde{\delta}_0.$
Finally, the approximate angular frequencies of the cavity are given by, 
\beq\label{s.23}
\begin{array}{l} 
\omega_{n,a}( \tilde{q}_0) = \frac{n\pi}{\xi_L}\left\{ 1 -  \frac{1}{2\xi_L}  \tilde{\delta}_0  (\alpha^2-1) \left[ 1-\cos ( 2n\pi \beta) \right] \right\}. \\
n=1,2,\dots.
\end{array}
\ene 

Note that the approximate angular frequencies (\ref{s.23}) depend on $\beta,$ i.e., they depend on the position of the membrane. On the contrary, the structural angular frequencies  given in (\ref{s.8}) and (\ref{s.9}) are independent of the position of the membrane. Notice that (\ref{s.8}) can be written as
 \beq\label{s23.b}
 \omega_{n,k}= \frac{n \pi}{\xi_L}+ \frac{\pi k}{\alpha \xi_L}, \quad n, k = 1,2\dots..
 \ene
 Keeping $k$ fixed and taking $n$ large we can make $\tilde{\delta}_0$ in (\ref{s.9})  small and, consequently, one can consider {\it structural frequencies} for a thin membrane. However, note that the  term $\frac{\pi k}{\alpha \xi_L}$ in (\ref{s23.b}) with $ k=1,2,\dots$ is not necessarily small. Hence, the structural frequencies in (\ref{s.8}) and (\ref{s.9}) may  not satisfy (\ref{s.21}) and (\ref{s.21b}). Moreover, introducing $\tilde{\delta}_0$ given by (\ref{s.9}) into (\ref{s.22}) we obtain
\beq\label{s.23c}
\begin{array}{l}
d_n= - \frac{1}{2\xi_L^2} \left\{ \frac{\xi_l k}{\alpha+ k/n}   \pi (\alpha^2-1) \left[1-\cos(2n \pi \beta) \right] \right\},
\end{array}
\ene
and we see that for the structural frequencies in (\ref{s.8}) and (\ref{s.9}) with $n$ large, the quantity $d_n$ is not necessarily small. 

 In Table~\ref{table} we compare the lowest twenty exact angular frequencies  $\omega_n(\tilde{q}_0)$ computed numerically using (\ref{s.7}) with the approximate angular frequencies   $\omega_{n,a}(\tilde{q}_0)$ evaluated using the analytic expression (\ref{s.23}). Observe that the approximation is very good, even  in the cases where the  upper bound in the error 
$\Delta_n$  in approximating the exact implicit equation (\ref{s.7}) by (\ref{s.19})  is of order one.  
\begin{centering}
\begin{table}[h]
\begin{tabular}{||c|c|c|c|c||}
\hline
$n$&$\omega_n$&$\omega_{n.a}$& $\Delta_n$ & \%\\
1&1.56241&1.56266&0.00293&-0.01548\\
\hline
2&3.09855&3.09897&0.01152&-0.01364\\
\hline
3&4.65208&4.64845&0.02597&0.07806\\
\hline
4&6.25509&6.25062&0.04695&0.07144\\
\hline
5&7.85374&7.85398&0.07402&-0.00309\\
\hline
6&9.36559&9.37594&0.10526&-0.11043\\
\hline
7&10.8399&10.8464&0.141&-0.05991\\
\hline
8&12.4205&12.3959&0.18512&0.19801\\
\hline
9&14.0847&14.0639&0.23805&0.14738\\
\hline
10&15.706&15.708&0.29601&-0.01245\\
\hline
11&17.1536&17.1892&0.3531&-0.20739\\
\hline
12&18.5839&18.5938&0.41443&-0.05359\\
\hline
13&20.2095&20.1433&0.49011&0.32742\\
\hline
14&21.9236&21.8772&0.57677&0.21167\\
\hline
15&23.5553&23.5619&0.66582&-0.02832\\
\hline
16&24.9283&25.0025&0.7457&-0.29766\\
\hline
17&26.3411&26.3412&0.83262&-0.00061\\
\hline
18&28.0175&27.8907&0.94198&0.45251\\
\hline
19&29.7693&29.6905&1.06346&0.26494\\
\hline
20&31.3999&31.4159&1.18314&-0.05103\\
\hline
\end{tabular}
\caption{\label{table}  We present the first twenty  angular frequencies, $\omega_n,$ computed numerically using the implicit equation (\ref{s.7}) and the analytic approximate angular frequencies $\omega_{n,a}$ given by 
(\ref{s.23}). The quantity $\Delta_n$ given in (\ref{s.23b}) is an upper bound of the error in approximating the exact implicit equation (\ref{s.7}) by the approximate implicit equation (\ref{s.19}).  In the column $ \%$ we give the percentage error between  $\omega_n$ and $\omega_{n,a}$, $100 \,(\omega_n-\omega_{n,a})/ \omega_n.$
The length of the cavity is $\xi_L=2$ the width of the membrane is $\tilde{\delta}_0=0.01,$  $ \beta=0.2$, the position of the left side  of the membrane is 
 $\tilde{q}_0- \frac{\tilde{\delta}_0}{2}= 0.4,$ and $\alpha:=\sqrt{1+4\pi\chi_0}=2.$ As the quantities in the columns 
 $\Delta_n$ and $\%$ are quite small, we have numerically computed all quantities $\omega_n, \omega_{n,a},\Delta_n,$ 
 and $ \%$ with 16 digits but we  quote them in the table with five significant decimals.}
\end{table}
\end{centering}

\subsection{Numerical computation of the potential }\label{numerical}

In this subsection we illustrate by a numerical experiment that the multiple scales approximation to the potential
 $\tilde{A}_0(\xi,\tau)$ can be very accurate, even if we only take in (\ref{A02})  the first  term that corresponds to the initial cavity mode  at $\tau=0$ and we disregard the second term that contains the contributions from all the cavity modes different from the initial cavity mode at $\tau=0.$ So, in this case the coherent coupling between different cavity modes due to the movement of the membrane is negligible.  We assume that at $\tau=0$ the potential is in the ground state, i.e., in the cavity mode $G_1[\xi, \tilde{q}(0)]$ that corresponds to the smallest angular frequency, $\omega_1[\tilde{q}(0)]$. In other words, in (\ref{13}) we take $N=1$ and, for simplicity, we also take $ g_{11}=0$. Hence, we have,
 
\begin{eqnarray} 
\label{n.1}
c_{1}(0) &=&1,  \quad\quad c_{m}(0) =0, \qquad m=2,\dots\cr
c_{m}'(0) &=&0, \qquad m=1, \dots.
\end{eqnarray}  
We denote by $ \tilde{A}_{0,1}^{(2)}(\xi , \tau)$ the following approximate potential, 

\begin{eqnarray}
\label{n.2}
\tilde{A}_{0,1}^{(2)}(\xi , \tau) &=&\frac{1}{2}  \alpha_{1} \left[ \tilde{q}(\tau) \right]  e^{-i t_{11}(\tau) } G_{1}[ \xi , \tilde{q}(\tau) ]  \left\{1+ i\tilde{q}'(\tau) \frac{\omega_{1}'\left[ \tilde{q}(\tau) \right]}{4 \omega_{1}\left[ \tilde{q}(\tau) \right]^{2}} \right\} \cr
&& +c.c. . 
\end{eqnarray}
Note that the approximate potential $\tilde{A}_{0,1}^{(2)}(\xi , \tau)$ corresponds to the two-term approximation 
$\tilde{A}_{0}^{(2)}(\xi , \tau)$ given in (\ref{A02}), where we have taken the initial conditions  (\ref{n.1}) and we have discarded the second term in (\ref{A02}) that corresponds to the contributions of all the cavity modes different from the initial cavity mode $G_{1}[ \xi , \tilde{q}(\tau) ].$

For  our numerical computations we choose the parameters of our MIM from recent experiments. In particular, \cite{MIM1} considers a MIM setup with  a cavity of length approximately equal to  $2 \lambda_0$ and a membrane made of silicon nitride. So, for the dimensionless length of our cavity we take $\xi_L=2$ and we consider a slab membrane with   $\alpha:=\sqrt{1+4\pi\chi_0}=2.$ Furthermore, we take the dimensionless width $ \tilde{\delta}_0= 1/100.$ 
 
 \textcolor{black}{We want to consider a motion of the membrane that gives rise to a periodic $\omega_{1}[\tilde{q}(\tau)]$ and to an $\tilde{A}_{0,1}^{(2)}(\xi , \tau)$ that is periodic in $\tau$. To this effect we take the following law for the movement of the membrane (other values of the constants are also possible):}
\beq\label{n.3}
\tilde{q}(\tau): = 1.2179272- 0.1 \cos( 0.01 \tau ).
  \ene
 This is an oscillation around the point  $1.2179272,$ slightly to the right of the center of the cavity with amplitude $0.1$ and angular frequency $0.01.$ \textcolor{black}{Note that a membrane that harmonically oscillates is a typical configuration in the dynamical Casimir effect \cite{Dodonov}.} The period of the oscillation of the membrane is  $200 \pi.$   Moreover,
 \beq\label{n.4} \vert \tilde{q}'(\tau)\vert \leq 10^{-3}, \qquad  \vert \tilde{q}''(\tau)\vert \leq 10^{-5}, \tau \in \mathbb R.
 \ene
\textcolor{black}{Observe from (\ref{n.2}) that the dependence of $\tilde{A}_{0,1}^{(2)}(\xi , \tau)$ on $\tau$ is through the functions $\tilde{q}(\tau)$, $\tilde{q}'(\tau)$, and $e^{-it_{11}(\tau)}$ where $t_{11}(\tau)$ is defined in (\ref{EscalaRapida}). From (\ref{n.3}) it follows that $\tilde{q}(\tau)$ and $\tilde{q}'(\tau)$ are periodic with period $200\pi$, so  $\tilde{A}_{0,1}^{(2)}(\xi , \tau)$ in (\ref{n.2}) will be periodic in $\tau$ with period $200\pi$ if $e^{-it_{11}(\tau)}$ is periodic with period $200\pi$. 
A numerical evaluation shows that
  \beq\label{n.4.b}
 t_{11}(200\pi)= 310 \pi= 155 (2\pi).
 \ene
 As the function $\tilde{q}(\tau)$ is periodic with period $200 \pi$ its integral over any interval of length $200 \pi$ is the same and equal to  $155 (2\pi).$ As a consequence, $e^{-it_{11}(\tau)}$ will be periodic in $\tau$ with period $200\pi$
 }
\textcolor{black}{Hence, the approximate potential $\tilde{A}_{0,1}^{(2)}(\xi , \tau)$ defined in (\ref{A02}) is periodic with period $ 200 \pi$.} We have chosen the numerical values of the coefficients in (\ref{n.3}) so that this periodicity holds. This is a particularly interesting case, since our numerical results show that the exact (numerical) potential  $\tilde{A}_{0}(\xi , \tau)$ approximately has this periodicity. 

We recall that the system of ordinary differential equations (\ref{10}) for the expansion coefficients $c_n(\tau)$ in (\ref{ExpansionEnModos}) is equivalent to the wave equation (\ref{8}). We find it more convenient to numerically solve the system (\ref{10})  than to solve the wave equation (\ref{8}). We have solved the system of differential equations (\ref{10}) using the software Mathematica. As we already mentioned below equation (\ref{A02}), the results of \cite{2016} imply that only the cavity modes that are near the initial cavity mode at $\tau=0$  give a significant contribution to the potential $\tilde{A}_{0}(\xi , \tau).$ As we take the first cavity mode as our initial cavity mode, only the first few modes will give a significant contribution. Naturally, to numerically solve the system of ordinary differential equations (\ref{10}) it is necessary to cut the series that appears on the right-hand side to a finite number of terms. By integrating (\ref{10}) several times and keeping different numbers of terms on right-hand side of (\ref{10}), we empirically determined that only the first four coefficients $c_n(\tau), n=1,2,3,4,$ are  nonnegligible and that it is enough to take ten terms in the series on the right-hand side of (\ref{10})  to determine   $c_n(\tau), n=1,2,3,4,$  in a stable way, so that the dependence on the number of terms in the series in negligible. So, summing up, in the numerical results on the potential $\tilde{A}_{0}(\xi , \tau)$ that we present below, we have taken the first four terms in the expansion of the potential given in (\ref{ExpansionEnModos}) and we have computed the coefficients $c_n(\tau), n=1,2,3,4$,  solving the system of ordinary differential equations (\ref{10}) keeping  ten terms in the series on the right-hand side. Including more terms  in  (\ref{ExpansionEnModos}) and/or more terms on the right-hand side in (\ref{10}) gives a negligible contribution to the numerical evaluation of the potential  $\tilde{A}_{0}(\xi , \tau). $
  
  To  quantify the difference between  the potential  $\tilde{A}_{0}(\xi , \tau)$ and the approximate potential
      $\tilde{A}_{0,1}^{(2)}(\xi , \tau)$ we introduce an appropriate norm. For this purpose, we denote by
       $L^2_{\tilde{q}(\tau)},
 \tilde{q}(\tau) \in [\tilde{\delta}_0/2, \xi_L-\tilde{\delta}_0/2],$ the  Hilbert space $L^2(0, \xi_L)$ endowed with the scalar product 
 $$
 (\phi,\psi)_{L^2_{\tilde{q}(\tau)}}:= \int_{0}^{\xi_{L}}d\xi \ \tilde{\epsilon}\left[ \xi - \tilde{q}(\tau) \right] \overline{\phi(\xi)} 
 \psi(\xi),
 $$
and we denote the associated norm by,
\beq\label{n.4.c}
 \|\phi\|_{L^2_{\tilde{q}(\tau)}}:= \sqrt{ (\phi,\phi)_{L^2_{\tilde{q}(\tau)}}}.
\ene
Note that the norms  $ \| \cdot \|_{L^2_{\tilde{q}(\tau)}}$ with $\tilde{q}(\tau) \in [\tilde{\delta}_0/2, \xi_L-\tilde{\delta}_0/2]$ are equivalent to each other.  We introduce the following convenient notation for $ \tau \in [0,\infty)$
\begin{eqnarray} \label{n.5}
a(\tau)&:=&  \|\tilde{A}_{0}(\cdot , \tau)\|_{L^2_{\tilde{q}(\tau)}},
\cr
b(\tau)&:=& \|\tilde{A}_{0,1}^{(2)}
(\cdot , \tau)\|_{L^2_{\tilde{q}(\tau)}},\cr
d(\tau)&:=& 100  \frac{ \|\tilde{A}_{0}(\cdot , \tau)-\tilde{A}_{0,1}^{(2)}
(\cdot , \tau)\|_{L^2_{\tilde{q}(\tau)}}}{  \|\tilde{A}_{0}(\cdot , \tau)\|_{L^2_{\tilde{q}(\tau)}}} .
\end{eqnarray} 
Note that $d(\tau)$ quantifies the difference between  $\tilde{A}_{0}(\xi , \tau)$ and $\tilde{A}_{0,1}^{(2)}(\xi,\tau)$
in the norm of $L^2_{\tilde{q}(\tau)}$ relative to   $\|\tilde{A}_{0}(\cdot , \tau)\|_{L^2_{\tilde{q}(\tau)}}$ and as a percentage.

 In  Table~\ref{table2}  we give the numerical values of $a(\tau), b(\tau),$ and $d(\tau)$ for several values of $\tau.$  Our numerical results  show that the  potential $\tilde{A}_{0}(\xi , \tau)$ given by (\ref{ExpansionEnModos}), where the expansion coefficients are solutions to the system of ordinary differential equations (\ref{10}) and  satisfy the initial conditions (\ref{n.1}), is very \textcolor{black}{accurately described by the two-term multiple scales solution} (\ref{n.2}) where we have discarded the contributions from all the cavity modes
different from the initial cavity mode, $G_{1}[ \xi , \tilde{q}(\tau) ]$, 
since the percentage error in $L^2$ norm is smaller or equal to $0.1481 \%.$
 This means that, to a good approximation, the cavity modes different from the initial one are not excited and that the  potential  $\tilde{A}_{0}(\xi , \tau)$ is well approximated by the simple analytical expression given in (\ref{n.2}). Hence, in this case the coherent coupling between different cavity modes due to the  movement of the membrane is negligible.

\begin{centering}
\begin{table}[H]
\begin{tabular}{||c|c|c|c||}
\hline
$\tau$&$a(\tau)$&$b(\tau)$& $d(\tau)$ \\
\hline
100&0.6358 &0.6359&0.05632  \\
\hline
200&0.3316 &0.3317&0.05558 \\
\hline
300&0.9990&0.9991&0.01233 \\
\hline
400&0.2781&0.2781&0.1418\\
\hline
500&0.7230& 0.7234&0.06772 \\
\hline
600&0.9894 & 0.9894 &0.002393 \\
\hline
200 $\pi$& 1  & 1     & 5$\times10^{-4}$\\
\hline
700&0.5294  &0.5293 &0.06134  \\
\hline
\end{tabular}
\caption{\label{table2}  We give the numerical values of the quantities $a(\tau), b(\tau),$ and $d(\tau)$ defined  in (\ref{n.5}) for $ \tau= 100, 200, 300, 400, 500, 600, 200\pi$, and $700$. Recall that $200\pi$  is the period in $\tau$ of the approximate potential $\tilde{A}_{0,1}^{(2)}(\xi,\tau)$ defined in (\ref{n.2}). Note that the quantity $d(\tau)$ given in the fourth column is extremely small. In particular, it is very small for the period, $\tau=200 \pi,$ of the approximate potential $\tilde{A}_{0,1}^{(2)}(\xi , \tau)$. This shows that the potential  $\tilde{A}_0(\xi,\tau)$ is well approximated by  $\tilde{A}_{0,1}^{(2)}(\xi , \tau)$ and that  $\tilde{A}_0(\xi,\tau)$ approximately has the same periodicity of the approximate potential   $\tilde{A}_{0,1}^{(2)}(\xi , \tau).$ We have taken the dimensionless length of the cavity $\xi_L=2,$ the dimensionless width of the membrane $\tilde{\delta}_0=1/100,$
 $\alpha:=\sqrt{1+4\pi\chi_0}=2,$ and the oscillation of the membrane as $\tilde{q}(\tau): = 1.2179272- 0.1 \cos(0.01 \tau).$}
\end{table}
\end{centering}

\begin{figure}[H]
\includegraphics[width=8cm]{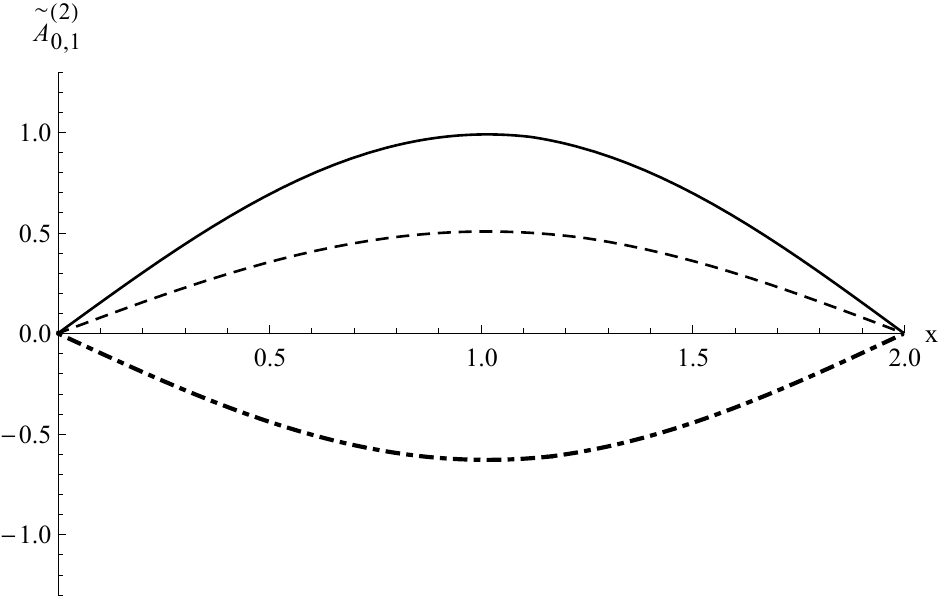}
\caption{\label{Figure3} We display the approximate potential  $\tilde{A}_{0,1}^{(2)}(\xi , \tau)$ for three values of the dimensionless time $\tau.$
The continuous line  corresponds to $\tau=0,$ the dotdashed line to $\tau= 100,$ and the dashed line to $\tau=190.$ Recall that   $\tilde{A}_{0,1}^{(2)}(\xi , \tau)$ is periodic in $\tau$  with period $200 \pi.$
We have taken the dimensionless length of the cavity $\xi_L=2,$ the dimensionless width of the membrane $\tilde{\delta}_0=1/100,$
$\alpha:=\sqrt{1+4\pi\chi_0}=2,$ and the oscillation of the membrane as $\tilde{q}(\tau): = 1.2179272- 0.1 \cos(0.01 \tau).$ }
\end{figure}

In Figure~\ref{Figure3} we display the approximate potential $\tilde{A}_{0,1}^{(2)}(\xi , \tau)$  for times $\tau=0, 100,190.$

\section{Conclusions}\label{conclusions}
We studied a {\it membrane-in-the-middle} optomechanical model  that consists of a one-dimensional  cavity with two perfectly conducting mirrors that are fixed, and a mobile, dielectric membrane inside the cavity. We assumed that the membrane is a linear, isotropic, non-magnetizable, non-conducting, and uncharged dielectric whose electric susceptibility is constant and that is allowed to move only along the axis of the cavity. As a consequence of the movement of the membrane,  the dynamics  of the cavity field was determined by a wave equation with time-dependent coefficients and modified by terms proportional to the velocity and acceleration of the membrane. We found a sequence of {\it structural angular frequencies} that have the following remarkable property: for each fixed value of the length of the cavity and each fixed value of the electric susceptibility of the membrane, there is a sequence of widths of the membrane such that, for each width, there there are cavity angular frequencies that are independent of the position of the membrane inside the cavity. These {\it structural  angular frequencies} are the only ones that have the remarkable property of being independent of the position of the membrane inside the cavity and are always larger than the smallest cavity angular frequency. It is noteworthy to point out that identifying and using the structural angular frequencies in experimental setups may lead to the study of new physics. 

\textcolor{black}{Although the {\it structural angular frequencies} are independent of the position of the membrane inside the cavity, the associated instantaneous cavity modes, of course, are not. As a consequence, this would lead to a cavity decay rate that depends on the position of the membrane when one considers the \textit{leakage of electromagnetic radiation out of the cavity.} This dissipative process could be modeled by a \textit{system + Markovian reservoir} approach \cite{Dutra} where the electromagnetic field inside the cavity could be quantized using the instantaneous cavity mode associated with a \textit{structural angular frequency}, and the method presented in \cite{Law} and the Markovian reservoir is composed of all the electromagnetic modes outside the cavity. This would lead to a Lindblad master equation where the cavity decay rate depends on the position of the membrane, that is, to an evolution equation where the dissipative coupling mentioned in the Introduction is present. Furthermore, as the  {\it structural angular frequencies} are independent of the position of the membrane, the dispersive coupling associated to the dependence of the angular frequency on the position of the membrane would be absent. This would allow the study of the dissipative coupling in the absence of the dispersive coupling mentioned in the Introduction.
 Then, the dissipative coupling could be used to perform tasks such as optomechanical cooling and mechanical sensing.}

\textcolor{black}{The \textit{membrane-in-the-middle} model  consisting of a one dimensional  cavity with a membrane that moves according to a law that is given by an external agent is one of the main  models in the study of the dynamical Casimir effect \cite{Dodonov}. Our results on {\it structural angular frequencies} can shed new light into the dynamical Casimir effect.  Our \textit{structural angular frequencies} can be used to study the dynamical Casimir effect in the new situation where there are  angular frequencies  independent of the position of the membrane.  Recall that even though the {\it structural angular frequencies} are independent of the position of the membrane inside the cavity, the associated instantaneous cavity modes are not. It would be interesting to study how the dynamical Casimir effect changes in this situation.
 A related problem is the Casimir effect with dynamical boundary conditions \cite{flm}, \cite{jw1}, \cite{jw2}.    In this case there is no membrane inside the cavity, but  at the boundary of the cavity  the boundary conditions are dynamical. The cavities with dynamical boundary conditions are  relevant for the realistic modelling  of the set up   of experiments verifying  the dynamical Casimir effect \cite{wj}. It is an interesting open problem to  study  our \textit{structural angular frequencies} for cavities  with dynamical boundary conditions, and to investigate how the dynamical Casimir effect changes in this case. }

In addition, we studied the case of a thin, slab membrane and found \textcolor{black}{simple, approximate formulae} for the angular frequencies and the modes of the cavity. Finally, we numerically computed the electromagnetic potential assuming that initially it is in the cavity mode that corresponds to the lowest cavity angular frequency. We took the parameters as in the experimental setup \cite{MIM1} and showed that the multiple scales approximation to the vector potential deduced in \cite{2016} is very accurate in this case.

\ack
This research was partially supported by project  PAPIIT-DGAPA UNAM IN100321. The numerical results where obtained using the Laboratorio Universitario de Alto Rendimiento, IIMAS-UNAM.
Ricardo Weder is an emeritus  fellow of the Sistema Nacional de Investigadores, CONAHCYT. This research was partially done while Ricardo Weder was visiting Instituto de F\'{i}sica Rosario, CONICET-UNR. He thanks Rodolfo Id Betan and Luis Pedro Lara for their kind hospitality.


\section*{References}
\begin{thebibliography}{99}

 \bibitem{Marquardt.2} Aspelmeyer M, Kippenberg T J  and  Marquardt F, Cavity optomechanics 2014 \textit{Rev. Mod. Phys.} \textbf{86} 1391
  
 \bibitem{Thompson}  Thompson J D,  Zwickl B M, Jayich A M, Marquardt F, Girvin S M, and  Harris J G 2008 \textit{Nature} \textbf{452} 7183 
  
 \bibitem{26} Kippenberg T J,  Rokhsari H,  Carmon T,  Scherer A, and Vahala K J 2005 \textit{Phys. Rev. Lett.} \textbf{95} 033901
  
 \bibitem{27} Carmon T, Cross M C, and Vahala K J 2007 \textit{Phys. Rev. Lett.} \textbf{98} 167203 

\bibitem{28} Metzger C, Ludwig M, Neuenhahn C, Ortlieb A, Favero I, Karrai K and Marquardt F 2008 \textit{Phys. Rev. Lett.} \textbf{101} 133903
  
\bibitem{39} Doolin C, Hauer B D, Kim P H,  Mac Donald A J R, Ramp H and  Davis J P 2014 \textit{Phys. Rev. A} \textbf{89} 053838
     
 \bibitem{29} Casta\~nos L O and Weder R 2014  \textit{Phys. Rev. A} \textbf{89} 063807

\bibitem{CS} Casta\~{n}os L O and  Weder R 2014 \textit{IOP Journal of Physics: Conference Series} \textbf{512} 012005 

\bibitem{PhysicaScripta}  Casta\~nos L O and R. Weder R 2015 \textit{Phys. Script.} \textbf{90} 068011
  
\bibitem{30} Bakemeier L and Alvermann A and Fehske H 2015 \textit{Phys. Rev. Lett.} \textbf{114} 013601
 
\bibitem{31} Krause A G,  Hill J T,  Ludwig M, Safavi-Naeini A H,  Chan J,  Marquardt F and  Painter O 2015 \textit{Phys. Rev. Lett.} \textbf{115} 233601

\bibitem{32} Buters F M,  Eerkens H J, Heeck K, Weaver M J,  Pepper B,  de Man S and Bouwmeester D 2015 \textit{Phys. Rev. A} \textbf{92} 013811 

\bibitem{33} Monifi F,  Zhang J,  \"{O}zdemir S K,  Peng B, Liu Y,  Bo F,  Nori F and Yang L 2016 \textit{Nat. Photon.}  
 \textbf{10} 399

\bibitem{34} Wang M,  L\"{u} X-Y,  Ma J-Y,  Xiong H, Si L-G and Wu Y 2016 \textit{Sci. Rep.} \textbf{6} 22705 

\bibitem{35} Schulz C, Alvermann A, Bakemeier L and Fehske H 2016 \textit{Europhys. Lett.} \textbf{113} 64002

\bibitem{Law} Cheung H K and Law C K 2011 \textit{Phys. Rev. A} \textbf{84} 023812 
  
\bibitem{2016} Casta\~nos L O and Weder R 2016 \textit{Phys. Rev. A} \textbf{94} 033846
  
 \bibitem{38} Leijssen R, Gala G R L,  Freisem L,  Muhonen J T and Verhagen E 2017 \textit{Nat. Commun.} \textbf{8} 16024  
 
 \bibitem{36} Wu J et al 2017 \textit{Nat. Commun.} \textbf{8} 15570

\bibitem{37} Navarro-Urrios D  et al 2017 \textit{Nat. Commun.} \textbf{8} 14965
  
 \bibitem{40} Djorwe P, Pennec Y, and Djafari-Rouhani B 2018 \textit{Phys. Rev. E} \textbf{98} 032201

\bibitem{Fig2020} Roque T F,  Marquardt F, and Yevtushenko O M 2020 \textit{New J. Phys.} \textbf{22} 013049
   
 \bibitem{CND} Hu C S,  Shen L T,  Yang Z B,  Wu H,  Li Y , and  Zheng S B 2019 \textit{Phys. Rev. A} \textbf{100} 043824
 
 \bibitem{gra1} Aso Y, Michimura Y, Somiya K, Ando M, Miyakawa O, Sekiguchi T,  Tatsumi D and Yamamoto H (KAGRA Collaboration) 2013 \textit{Phys. Rev. D} \textbf{ 88} 043007
 
 \bibitem{gra2} Abbott B P  et al (LIGO Scientific Collaboration and Virgo Collaboration) 2016 \textit{Phys. Rev. Lett.} \textbf{116} 061102
  
 \bibitem{Converter}  Higginbotham A P,  Burns P S,  Urmey M D, Peterson R W,  Kampel N S,  Brubaker B M,  Smith G,  Lehnert K W  and Regal C A 2018 \textit{Nat. Phys.} \textbf{14} 1038
  
 \bibitem{MIM2} Rodrigues I,  Bothner D and Steele G 2019 \textit{Nat. Commun.} \textbf{10} 5359 
  
 \bibitem{Room2} Faust T, Krenn P,  Manus S, Kotthaus J and Weig E 2012 \textit{Nat. Commun.} \textbf{3} 728
  
 \bibitem{Room} Le A T,  Brieussel A and Weig E M 2021 \textit{J. App. Phys.} \textbf{130} 014301
  
 \bibitem{circuit} Zhou X Cattiaux D, Theron D and Collin E 2021 \textit{J. App. Phys.} \textbf{129} 114502

\bibitem{MST} Golokolenov I, Cattiaux D, Kumar S, Sillanp\"a\"a M,  de L\'epinay L M.,  Fefferman A and Collin E 2021 \textit{New J. Phys.} \textbf{23} 053008

\bibitem{QCP} Armata F,  Lamiral L, Pikovski I, Vamer M R,  Brukner \v{C} and Kim M S 2016 \textit{Phys. Rev. A} \textbf{93}  063862
  
\bibitem{Li2019} Li X,  Korobko M,  Ma Y,  Schnabel R and  Chen Y 2019 \textit{Phys. Rev. A} \textbf{100} 053855  

\bibitem{MOUT} Tagantsev A K and Polzik E S 2021 \textit{Phys. Rev. A} \textbf{103} 063503
 
\bibitem{MIM1}  Hornig G J,   Al-Sumaidae S,  Maldaner J,  Bu L and  Decorby R G 2020 \textit{Opt. Express} \textbf{28} 28113

\bibitem{QDC2} Dummont V, Lau H-K, Clerk A A and Sankey J C 2022 \textit{Phys. Rev. Lett.} \textbf{129}  063604

\bibitem{Dodonov} Dodonov V V 2020 \textit{MDPI Physics} \textbf{2} 67 

\bibitem{Sobolev} Adams R A and Fournier J J F 2003 \textit{Sobolev Spaces} (2nd ed., Academic Press, New York)

\bibitem{rs} Reed M and Simon B 1978 \textit{Methods of Modern Mathematical Physics IV Analysis of Operators} (Academic Press, New York)

\bibitem{Dutra} Dutra S. M. 2005 \textit{Cavity Quantum Electrodynamics: The Strange Theory of Light in a Box} (Wiley, New Jersey)


\bibitem{flm}Fosco  C~ D,  Lombardo  F.~C  and   Mazzitelli F~D 2013 \textit { Phys. Rev. D}  \textbf{87} 105008

\bibitem{jw1}  Ju\'arez-Aubry B~A, Weder R 2021 \textit{J. Phys. A : Math Theor} \textbf{54} 105203

\bibitem{jw2} Ju\'arez-Aubry B~A, Weder R 2023  \textit{Quantum field theory with dynamical boundary conditions and the Casimir effect.} In \textit{Theoretical Physics, Wavelets, Analysis, Genomics An indisciplinary Tribute to Alex Grossman}, Ed Flandrin P, Jaffard S, Thierry P and Torr\'esani B, (Birkh\"auser Switzerland), pp 195-218  

\bibitem{wj}   Wilson C~M , Johansson G , Pourkabirian A, Simoen  M, Johansson J.~R. ,
 Duty  T,  Nori F, and Delsing  P  2011  \textit{Nature} \textbf{479} 376


\end{thebibliography}
\end{document}